\documentclass[aps,pre,twocolumn,showpacs]{revtex4}

\usepackage{epsfig}
\usepackage{amsmath}
\usepackage{amssymb} 
\usepackage{graphicx}
\usepackage{graphics}
 
\usepackage{color}

\begin{document}
\newcommand{\tl}{\tilde{l}}

\title{Monodisperse domains by proteolytic control of the coarsening instability}
\author{Julien Derr$^{1,2,3}$}
\email{julien.derr@univ-paris-diderot.fr}

\author{Andrew D. Rutenberg$^1$}
\email{andrew.rutenberg@dal.ca}
\affiliation{$^1$Department of Physics and Atmospheric Science, Dalhousie University, Halifax, Nova Scotia, Canada, B3H 3J5 
	\\$^2$ FAS Center for Systems Biology, Harvard University, Northwest Labs, 52 Oxford Street, Cambridge, MA 02138, USA
	\\$^3$ Laboratoire Mati\`ere et Syst\`emes Complexes, Universit\'e Paris Diderot, 10 rue Alice Domont et L\'eonie Duquet, 75205 Paris cedex 13, France}
\date{\today}

\begin{abstract}
The coarsening instability typically disrupts steady-state cluster-size distributions. We show that degradation coupled to the cluster size, such as arising from biological proteolysis, leads to a novel fixed-point cluster size. Stochastic evaporative and condensative fluxes determine the width of the fixed-point size distribution. At the fixed-point, we show how the peak size and width depend on number, interactions, and proteolytic rate.  This proteolytic size-control mechanism is consistent with the phenomenology of pseudo-pilus length control in the general secretion pathway of bacteria.
\end{abstract}
\maketitle

\section{Introduction}
Living cells control the sizes of subcellular structures. Mechanisms of size control include molecular rulers for bacterial injectisome length \cite{Cornelis2003}, measuring cups for the length of the bacterial flagellar hook \cite{Makishima2001}, counting for telomere ends \cite{Markand1997}, and equilibrium energetics for actin bundle radius \cite{bundle}.  Size control is also evident in the length of eukaryotic flagella \cite{Rosenbaum2003} or the size of lipid rafts \cite{Pralle2000}. 

Size control is challenging in bacteria because of the strong stochastic effects expected in such small cells.  It is especially interesting how bacteria control the size of extracellular macromolecular assemblies, such as bacterial secretion systems and pili. In this paper, we investigate a novel length-control mechanism that may apply to the pseudopilus (here ``ppilus'') of the type-II secretion system (T2SS) of Gram-negative bacteria \cite{T2SS}. In the general secretory pathway, proteins are first secreted across the bacterial inner membrane by the Sec or Tat systems then across the outer membrane by the T2SS. The T2SS pushes folded proteins across the periplasm, and out a secreton in the outer membrane, using an assembling and disassembling ppilus that is thought to function as a piston or plunger \cite{T2SS}.  The ppilus assembles from the energized inner membrane and spans the periplasmic space.  The primary pilin subunit that assembles into the pseudopilus of the T2SS is variously called PulG in {\em Klebsiella oxytoca}, XcpT in {\em Pseudomonas aeruginosa}, or more generally GspG (here ``G'') and is homologous to the PilA pilin of the type-IV pilus used in twitching motility in, e.g., {\em P. aeruginosa} or {\em Myxococcus xanthus} \cite{Jarrell2008}.

For a functioning T2SS the ppilus length should span the periplasm, which is approximately 21 nm across \cite{Matias2003} -- or 85 G monomers in the ppilus structure \cite{Kohler2004}. The ppilus is not normally seen outside the cell \cite{T2SS}, hence the ``pseudo'' prefix, indicating an effective length-control mechanism. Overexpression of G leads to visible extracellular ppili \cite{Durand2005,Sauvonnet2000,Vignon2003} which rules out  the (fixed-size) molecular ruler or measuring cup mechanisms of size regulation.  

Underexpression of a minor pilin (XcpX \cite{Durand2005}, PulK \cite{Vignon2003}, or GspK ---  here ``K'') leads to long extracellular ppili. So, it has been proposed that simple stoichiometric control applies to T2SS ppilus length control \cite{Minamino2005}.  However, stoichiometric control faces the inherent challenges of precisely controlled protein expression \cite{UriAlon}.  Furthermore, while a single  ``stoichiometric'' ppilus with a fixed size pool of G monomers can have a narrow length distribution peaked at the pool size \cite{onestoichiometric}, this is not true of multiple ppili sharing a common pool of G monomers.  With multiple ppili stoichiometric length distributions are exponential, as seen in stochastic simulations with more than one ppili (see Appendix A).   Since five to ten ppili are present on each individual bacteria \cite{Buddelmeijer2006}, stoichiometric length control of each individual  ppilus requires an additional mechanism to partition G proteins equally between the ppili.

\section{Model}
The observations of G-G interactions \cite{Pugsley1996}  and of G-clusters in individual bacteria \cite{Pugsley1996,Scott2001}  are consistent with clusters of G that could be associated with each ppilus, as illustrated by the dashed regions in Fig.~1.  The size of each cluster, i.e. the number of G monomers, would then determine the maximal length of the associated ppilus, converting a ppilus length-control problem into a G-cluster size-control problem.  Nevertheless, thermally driven evaporation, condensation, and diffusion (Fig.~1 (e), (c), and (D) respectively) will destabilize spontaneous partitioning of G among many clusters. The subsequent coarsening of the size-distribution, treated by Lifshitz, Slyozov and Wagner (LSW) \cite{Bray2002}, would lead to a single large cluster (condensed phase) of G in equilibrium with small clusters of G (vapour) in the bacterial inner membrane.  The net growth of large diffusively-coupled $2d$ clusters can be expressed in terms of their radius $R$ by $dR/dt = A \left( 1/R_c - 1/R \right)/R$, where $A$ is a constant.  Cluster growth is the net result of a condensation flux (due to the supersaturation of G monomers in the inner membrane) and an evaporative flux (due to the curved boundary of the cluster within the membrane).  $R_c$ is the critical cluster size above which clusters grow and below which they shrink, and it typically grows with time due to the decreasing supersaturation associated with increasing average cluster radius via the Gibbs-Thompson effect. Growth of the average cluster size is also associated with a decreasing number of clusters, and with a broad distribution of cluster sizes \cite{Bray2002}. This is qualitatively unchanged for collections of smaller or irregularly-shaped clusters.  

\begin{figure} \begin{center}
 \includegraphics[width=3.25in]{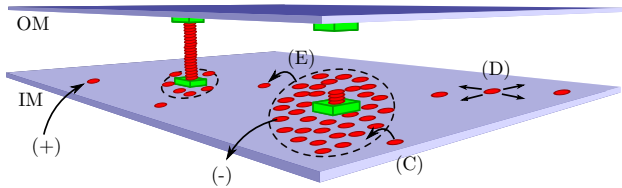}
\caption{(Color online) Cartoon of the bacterial periplasmic space showing the inner (IM) and outer (OM) membrane, with pilin ``G'' proteins (red circles). Two pseudo-pili are shown, extending from their inner membrane base towards their outer membrane secretons (green squares). 
The IM clusters associated with the ppili are indicated by the dashed circles. The number of G proteins in the ppilus together with its associated cluster is unchanged by ppilus assembly or disassembly (not shown).  Numbers change by the physical processes of condensation (c), evaporation (e), proteolysis or recycling (-), and insertion (+).  Diffusion (D) of monomers couples the clusters. }
\label{fig:cartoon} 
\end{center} 
\end{figure}

\begin{figure} \begin{center}
 \includegraphics[width=\linewidth]{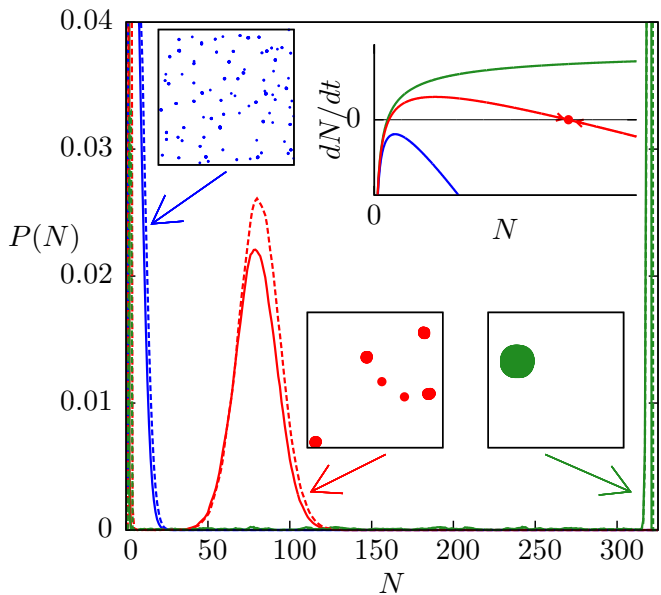}
\caption{(Color online) Cluster size probability distribution, $P(N)$, vs. number of monomers in a cluster, $N$, for low ($\alpha=10^{-9}$, green, right), medium ($\alpha= 5 \times 10^{-8}$, red, center), and high ($\alpha=10^{-5}$, blue, left) proteolytic rates (per protein per timestep, where $\Delta t \approx 10^{-4}s$). Other parameters are $\tilde{J}=1.67$ and $\rho=2 \times 10^{-3}$. The dotted lines represent the reconstruction of the probability distribution via Eqn.~\protect\ref{PofN}. Intermediate proteolytic rates stabilize a peaked steady-state cluster size distribution with multiple large clusters, as illustrated by the corresponding snapshot of the system. The inset is a stability diagram from Eqn.~\protect\ref{Adot}, illustrating the effects of low (green, top curve, with only an unstable fixed point), medium (red, center curve,  with an additional stable fixed point as indicated), and high (blue, bottom curve, with no fixed points) proteolytic rates on the domain size dynamics $dN/dt$ vs. domain size $N$.}
\label{fig:snaps} 
\end{center} 
\end{figure}

In a biological system we should add both protein synthesis and proteolysis.  Protein synthesis will simply contribute to the supersaturation, but proteolysis will add a new degradative term to the LSW dynamics proportional to the number of monomers in the cluster.   Bacterial proteolytic mechanisms include cytoplasmic ATP-dependent proteosomes \cite{Butler2006}, ubiquitin-like targeting systems \cite{Darwin2009}, and periplasmic ATP-independent proteases \cite{Huber2008}.  Non-degradative recycling of components away from the membrane has an equivalent effect \cite{Gomez,Turner2005,Fan2008}. In terms of the number of monomers in a large $2d$ cluster of size $N$ we will then have 
\begin{equation}
 	\frac{dN}{dt} = S  - \frac{E}{\sqrt{N}} - \alpha N,
	 \label{Adot}
\end{equation}
where  $S$ corresponds to condensation due to supersaturation, $E$ corresponds to evaporation, and $\alpha$ is the proteolytic rate. Eqn.~\ref{Adot} is an approximate mean-field equation, without stochastic effects or higher-order curvature corrections expected for smaller clusters. However, it illustrates  (see inset of Fig.~\ref{fig:snaps}) how intermediate values of the  proteolytic term should generically stabilize coarsening to a steady-state cluster size (red dot in inset), while for large enough proteolytic rate no stable fixed-point exists (blue curve in inset).

To test these ideas we model G clustering in the bacterial membrane with a stochastic Ising lattice-gas in two-dimensions (see e.g. \cite{Ryan}). The dynamics are (conserved) particle-exchange, subject to a Metropolis acceptance criterion with a reduced interaction energy $\tilde{J} \equiv J/(k_B T)$. We supplement these dynamics with a dimensionless proteolysis rate $\alpha$ (per monomer per timestep), where (to minimize finite-size effects) monomers are removed and replaced at random positions in the system. Using a typical monomer size of $\Delta x =5nm$ and diffusivity of $D= 70000 nm^2/s = \delta x^2/(4 \delta t)$ we have $\Delta t \approx 10^{-4} s$ \cite{Ryan}, and the proteolytic lifetime is $\tau = \Delta t/\alpha$.  In units of $\Delta x$, we use a linear lattice size $L=400$ and check that finite-size effects are not significant.  A typical bacterium is larger in every direction, with $L \approx 2000$ \cite{Ryan}. 

\section{Results}
Figure~\ref{fig:snaps} illustrates the steady-state results of the stochastic simulations.  For a high proteolytic rate ($\alpha=10^{-5}$, or protein lifetime $\tau \approx 10 s$) no large domains are seen. At an intermediate proteolytic rate ($\alpha=5 \times 10^{-8}$, or $\tau \approx 2000s$), stable domains with a characteristic size are seen. At a low proteolytic rate ($\alpha=10^{-9}$, or $\tau \approx 10^{5}s$) proteolysis is insufficient to stabilize the largest domain against coarsening to the limits of the system size. For the remainder of the paper, and for intermediate proteolytic rates, we quantify the peak size $N_0$ as well as the full-width-half-maximum $W$ of the steady-state distribution of domain sizes, and explore how they vary with proteolytic rate, total membrane density of G, and reduced interaction $\tilde{J}$. 
 
The growth dynamics of a given cluster of size $N$ is given by the difference of incoming and outgoing flux,  $\dot{N} = \dot{N}_{+}(N) - \dot{N}_{-}(N)$. This gives us the transition probabilities of monomer addition $ \Gamma_{+}(N) \equiv \dot{N}_+/(\dot{N}_-+\dot{N}_+) $ and subtraction $\Gamma_{-}(N) \equiv  \dot{N}_-/(\dot{N}_-+\dot{N}_+)$. If $P(N)$ is the resulting steady state probability distribution of clusters of size $N$, then the detailed balance condition of the transition probabilities is $\Gamma_+(N-1)P(N-1) = \Gamma_-(N)P(N)$.  Approximating $P(N)$ as a continuous distribution of cluster sizes (where $dP/dN \simeq P(N)-P(N-1)$), then in steady state we have
\begin{equation}
	P(N) \simeq P(1) \exp{ \left[ \int_{1}^N \! \left( 1-\frac{\Gamma_-(n)}{\Gamma_+(n-1)}\right) \, dn \right]},
	\label{PofN}
\end{equation}
where we choose $P(1)$ such that $\sum_m P(m)=1$.

\begin{figure*}\begin{center} 
\includegraphics[width=0.33\linewidth,angle=-90]{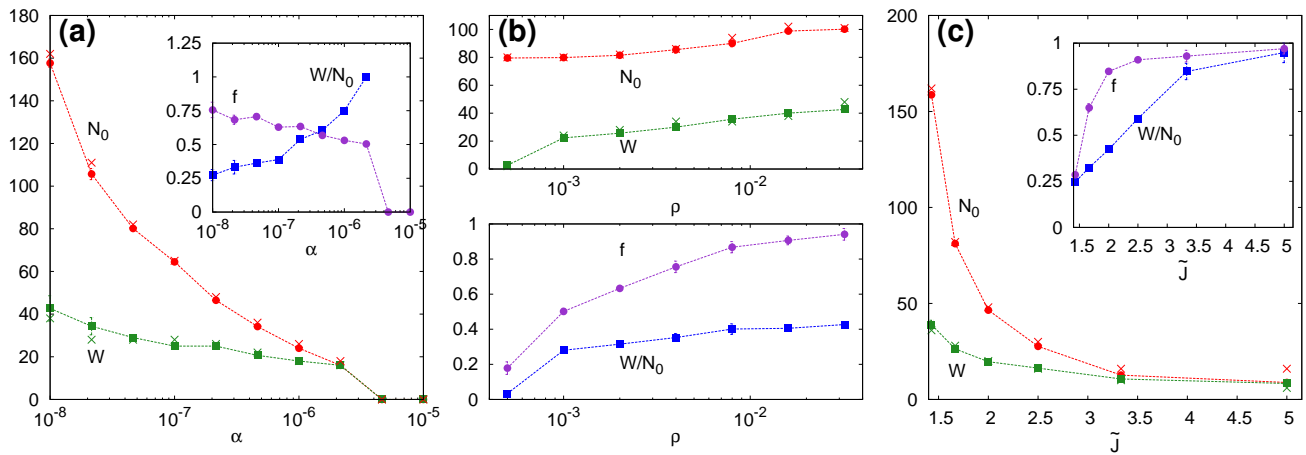}
\caption{(Color online)  (a) Dependence of the steady-state most-likely cluster size $N_0$ (red circles) and the full-width-half-maximum $W$ (green squares) of the cluster-size distribution vs. the proteolytic rate $\alpha$. Crosses ($\times$) indicate corresponding values derived from fluxes near $N_0$ via Eqn.~\ref{PofN}. The inset shows the relative width $W/N_0$ (blue squares) and the fraction of G-proteins in large domains (purple circles) vs. $\alpha$. Error bars indicate statistical errors, which are often smaller than point sizes. Other parameters are $\tilde{J}=1.67$ and $\rho=2 \times 10^{-3}$.  (b) The top figure shows size $N_0$ (red circles) and width $W$ (green squares) vs.  average membrane density $\rho \equiv N_g/L^2$.   The bottom figure shows the dependence of relative width $W/N_0$ (blue squares) and the cluster fraction $f$ (purple circles). Other parameters: $\tilde{J}=1.67$ and $\alpha=5\times10^{-8}$.  (c) Size $N_0$ (red circles) and width $W$ (green squares) vs the clustering interaction $\tilde{J} \equiv J/(k_B T)$.  The inset shows the relative width $W/N_0$ (blue squares) and the fraction of G-proteins in large domains $f$ (purple circles) vs. $\tilde{J}$. Other parameters: $\rho=2 \times 10^{-3}$ and  $\alpha=5\times10^{-8}$.}
\label{fig234} 
\end{center} \end{figure*}

We have measured the steady-state evaporation and condensation rates $N_\pm(N)$, and used the resulting transition probabilities $\Gamma_\pm(N)$ in Eqn.~\ref{PofN} to compare with the measured $P(N)$.  We find reasonable agreement (see e.g. the dotted lines in Fig.~\ref{fig:snaps} or the crosses in Fig.~\ref{fig234}), showing that the dispersity of domain sizes around the stable fixed point $N_0$ (where $N_+(N_0) = N_-(N_0)$) is due to stochastic fluctuations driven by the finite fluxes.  Intuitively, and as proven in Appendix B, the stochastic width $W$ of the size-distribution increases either with increasing absolute flux at the fixed point (i.e. $|N_\pm|$) {\em or} with decreasing net (stabilizing) flux ($|N_+ - N_-|$)  near the stable fixed point.  
  
Fig.~\ref{fig234}(a) shows the peak position and the width of the steady-state cluster-size distribution vs. proteolytic rate $\alpha$.  Increasing $\alpha$ decreases the stable domain size until, for $\alpha \gtrsim 4 \times 10^{-6}$, the distribution becomes peaked at $N_0=0$.  This is consistent with Eqn.~\ref{Adot} and the stability plots in the inset of  Fig.~\ref{fig:snaps}. The width decreases with $\alpha$ as the net stabilizing flux near the stable point increases.   As shown by the inset of Fig.~\ref{fig234}(a), the relative width $W/N_0$ nevertheless increases with $\alpha$. The fraction $f$ of G-monomers that are found in larger clusters (under the second peak in Fig.~2) slightly decreases with $\alpha$ but remains a considerable fraction (more than half) of the total, which indicates that proteolysis can be an efficient size-control mechanism.  Experimentally, it appears that at least $20\%$ of G-monomers are in ppili \cite{Hu2002}.  
  
In contrast, the steady-state peak cluster size $N_0$, the width $W$ near the peak, and the fraction of G in the peak clusters stay roughly constant over a large range of average G densities, $\rho \equiv N_g/L^2$, as shown in Fig.~\ref{fig234}(b).   We understand this as an effective (nonequilibrium) coexistence between a fixed density of G monomers and excess G in clusters ---  increasing $\rho$ simply increases the number of clusters without significantly changing their size-distribution.  (When approximately one cluster is seen in the system, for smaller $\rho$, finite-size effects do appear --- decreasing both $N_0$ and $W$. This is beginning to be apparent at the lowest $\rho$ in Fig.~\ref{fig234}(b). Strong finite-size effects are seen in the condensed-cluster fraction $f$, but simply arise from an approximately constant vapour density as $\rho$ varies.) The lack of strong dependence of the cluster size $N_0$ on the total membrane density $\rho$  is advantageous in terms of robust control of cluster size in the face of stochastic protein expression. 
 
At a fixed proteolytic rate and expression level, the effects of varying $J/(k_B T)$ (i.e. G-G interactions) are shown in Fig.~\ref{fig234}(c).  Thermal evaporation decreases with increasing $\tilde{J}$ --- leading both to an increasing $f$ (see inset of Fig.~\ref{fig234}(c)), a decreasing effective supersaturation, and a smaller stable $N_0$.   The fractional width $W/N_0$ of the peak of the cluster size distribution is narrower for weaker interactions, but at the same time a smaller fraction of G-monomers are in clusters.

\section{Discussion}
We find that intermediate levels of proteolysis controls the natural coarsening instability of condensed clusters, and leads to steady-state clusters with a relatively narrow distribution of sizes. For an intermediate proteolytic rate $\alpha=5 \times 10^{-8}$ (turnover time $\tau \sim 2000$s), we obtain a cluster size $N_0 \approx 80$ --- remarkably close to the $85$ G required to assemble a ppilus that spans the periplasmic space \cite{Matias2003,Kohler2004}. The fractional width is then about $30 \%$, consistent with the lack of extracellular ppili observed under normal conditions.  For cluster size-control to effectively control ppilus length, we predict that G-clusters are associated with the secreton base of the T2SS.   Mutational variation of G-monomers to affect either their proteolytic susceptibility ($\alpha$, see e.g. Fig.~\ref{fig234}(a)) or G-G interactions ($\tilde{J}$, see Fig.~\ref{fig234}(c)) should also affect the distribution of ppilus lengths.   We believe stoichiometric mechanisms, via GspK  \cite{Durand2005,Sauvonnet2000,Vignon2003}, control the ppilus length from individual associated G-clusters, while our proteolytic size control mechanism ensures that multiple G-clusters remain approximately equally sized.  

We have shown that a novel mean-field fixed-point in the cluster size distribution arises from proteolysis, while the dispersity of cluster sizes around the fixed point arises from stochastic growth and shrinkage of clusters. The noise associated with proteolysis is intrinsically multiplicative, in that proteolysis only targets existing G proteins.  Our lattice-gas model naturally implements both proteolysis and the thermal evaporation and condensation of clusters in a membrane.   

The result is a monodisperse cluster-size distribution with a non-zero peak size, which qualitatively differs from some earlier work on lipid raft sizes in membranes that only found distributions peaked at $N_0 \approx 0$ \cite{Turner2005,Fan2008}. We believe this is due to the approximate evaporation/condensation dynamics \cite{Turner2005} or the additive noise \cite{Fan2008} used in those works.  In contrast, earlier coarse-grained models of ternary mixtures with recycling \cite{Gomez} --- also applied to lipid nanodomains --- did recover a non-zero peak size, though did not include recycling noise.  Our microscopic two-component model is simpler, and the mean-field flow we present shows how the results are expected to be generic for proteolysis or recycling. 

Proteolysis is not just for cellular cleanup. Targeted degradation can adjust timescales and levels of transcription or translation (see, e.g. \cite{UriAlon}).  We have shown how it can also be used to control cluster sizes within the cell. Proteolysis contributes a new ``evaporative'' term that limits coarsening with a novel mean-field fixed point for the cluster size (inset of Fig.~\ref{fig:snaps}).  The same mechanism will qualitatively apply whether the proteolysis targets all proteins in a cluster (as in Eqn.~\ref{Adot}), appropriate for cytoplasmic or periplasmic proteases --- or whether it targets the cluster periphery as might be appropriate for membrane associated proteases. In both cases the loss term in Eqn.~\ref{Adot} will grow with $N$, and so will lead to a stable fixed point at some $N_0$. Proteolysis provides a new size-control mechanism to cells.  We expect that proteolysis or analogous degradation terms, such as recycling, are widely exploited to achieve monodisperse steady-state clusters in other biological systems.

\section*{Acknowledgments}
This work was supported by Natural Sciences and Engineering Research Council (NSERC), Canadian Institutes for Health Research (CIHR), and Atlantic Computational Excellence Network (ACENET); computational resources came from ACENET. J.D. was also supported by the Human Frontier Science Program. We thank Patrick McKelvey for stochastic simulations of the stoichiometric length-control mechanism. We acknowledge useful discussions with Olivera Francetic and Anthony Pugsley.


\section*{Appendix A - Stoichiometric length-control}
We present here analytical limits as well as computational simulations for a stoichiometric model for length-control. This  system consists of $M$ polymerized pili sharing a common pool of $N_g=MN_{p}$ monomers, where $N_{p}$ is the average number of monomers per pilus. Each pilus has a probability $p_+$ of growing and $p_-$ of terminating growth and completely disassembling, and these probabilities are dependent only on the pool of monomers. The total number of monomers is constant, so that $N_g=n+\sum_{i=1}^{M}l_i$, where $l_i$ are the lengths of the pili and $n$ the number of monomers remaining in the monomeric pool. Since the growth rate is proportional to $n$, while the termination rate $t$ is fixed, then the growth and termination probabilities are $p_+(n)=n/(n+t)$ and $p_-(n)=t/(n+t)$, respectively, where $p_++p_-=1$.

\begin{figure} \begin{center} 
\includegraphics[width=\linewidth]{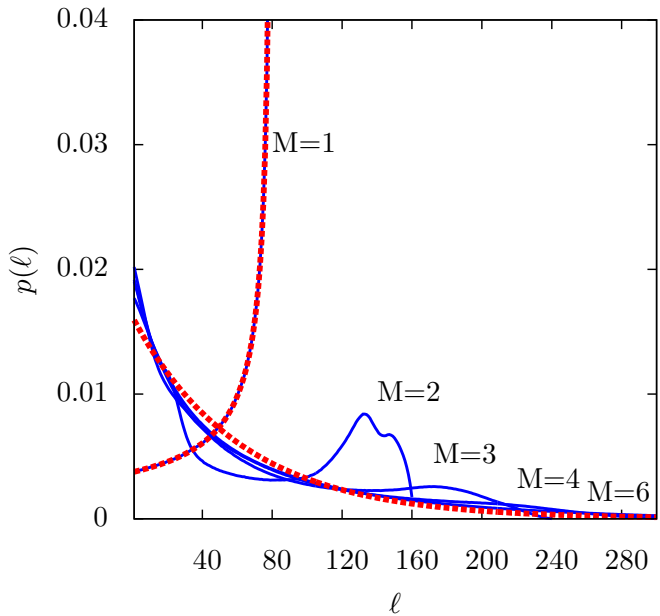}
\caption{(Color online) Probability distribution $p_M(l)$ of the maximum length for various number of pili ($M=1, 2, 3, 4, 6$) sharing the same common pool of monomers. We use $t=0.3$ and $N_{p}=40$, though qualitatively similar results are seen with other parameters. The continuous lines are the result of the stochastic simulation. The red dashed lines represents the two exact limiting distributions $p_1$ and $p_{\infty}$, overlaying the $M=1$ data and close to the $M=6$ data, respectively.}
\label{multiplepili} 
\end{center} \end{figure}

For a single pilus, the probability of achieving a maximum length $l$ can be written as the product of the individual probabilities: $p_1(l)=p_+(N_g) \times p_+(N_g-1) \times  \dots \times p_+(N_g-(l-1)) \times  p_-(N_g-l)$. This leads to 
\begin{equation}
	p_1(l)=\frac{t \Gamma(N_g +1) \Gamma(N_g +t -l)}{\Gamma(N_g +t +1) \Gamma(N_g -l +1)}
\label{limit1}
\end{equation}
which is peaked around $N_p$ for $t \lesssim 1$, as shown with $M=1$ in Figure~\ref{multiplepili}.  A single pilus can achieve length-control by assembling most of the available pool of monomers before disassembly is triggered.

For the more biologically appropriate case of multiple pili per cell, we can use a mean-field approximation.  In this case, $t$ becomes $Mt$, and $N_g$ becomes $MN_{p}-M\tl$ where $\tl$ is the average length of the pili, and then $p_M(l)=p_+(MN_{p}-M\tl) \times p_+(MN_{p}-M\tl-1) \times \dots \times  p_+(MN_{p}-M\tl-(l-1))\times  p_-(MN_{p}-M\tl-l)$. The $p_+$ are constant at $O(1/M)$, so that in the limit $M \gg 1$ we recover an exponential distribution
\begin{equation}
	p_{\infty}(l)=\frac{t+1}{N_{p}+t}e^{-\frac{t+1}{N_{p}+t} l}
\label{limit2}
\end{equation}

For the intermediate regime, with finite $M>1$, we performed stochastic computer simulations --- as illustrated in Fig.~\ref{multiplepili}. The exponential limit is quickly approached for $M \gtrsim 4$. However, even for $M=2$ the peak around $N_p$ seen for $M=1$ is lost.

\section*{Appendix B - Analytical monotonicities of the width}  

\begin{figure}[h] \begin{center}
\includegraphics[width=0.75\linewidth]{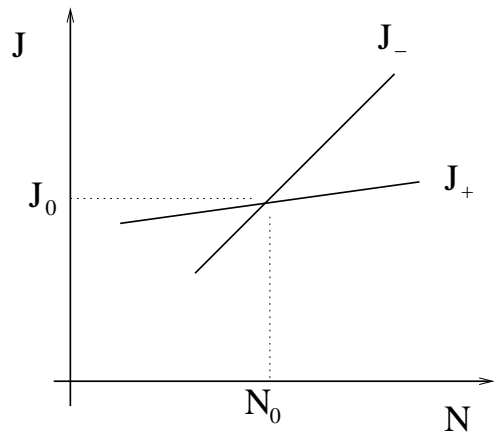}
\caption{Figure illustrating the mean-field fixed-point cluster size $N_0$ that occurs where the evaporative and condensative fluxes are equal ($J_-=J_+=J_0$). The fixed-point is stable when evaporation is stronger than condensation ($J_->J_+$) for $N >N_0$ --- as illustrated. The demonstration of monotonicities in this appendix are for $N \approx N_0$, where a linear approximation for $J_\pm$ holds.}
\label{linearfluxes}
\end{center} \end{figure}

For the proteolytic size-control mechanism described in the text, we detail here the formal derivations of the variation of the full width half maximum (FWHM) $W$ of the non-zero peak in the size distribution. We consider the evaporative and condensation fluxes $J_+$ and $J_-$, respectively, close to the stable fixed point size $N=N_0$, where $J_+=J_-=J_0$ (see Fig.~\ref{linearfluxes}). We allow for linear dependence of the fluxes near the fixed point
\begin{eqnarray}
	J_+ &=& a(N-N_0)+J_0 \\
	J_- &=& b(N-N_0)+J_0
\end{eqnarray}
where the stability of the fixed point requires $J_- > J_+$ for $N>N_0$ and $J_+> J_-$ for $N<N_0$, corresponding to the requirement that $b<a$. 

These fluxes give us the transition rates
\begin{eqnarray}
	\Gamma_+ &=& \frac{an+J_0}{cn+2J_0} \\
	\Gamma_-  &=& \frac{bn+J_0}{cn+2J_0},
\end{eqnarray}
where we define $n \equiv N-N_0$ and $c\equiv a+b$. The detailed balance condition is $p(n)=p(n-1) \Gamma_+(n-1)/\Gamma_-(n)$. We define
\begin{equation}
\begin{split}
	A(n) \equiv & \displaystyle\prod_{m=1}^n  \frac{\Gamma_+(m-1)}{\Gamma_-(m)} \\
                    &  = \frac{cn/2+J_0}{bn+J_0} \frac{\prod_{m=1}^{n-1}(ma+J_0)}{\prod_{m=1}^{n-1}(mb+J_0)}.\\
\end{split}
\label{condition}
\end{equation}
$A$ monotonically decreases with $n$, since $\Gamma_+(n-1)/\Gamma_-(n) = (a(n-1)+J_0)/(b(n-1)+J_0)<1$ where $b>a$.  The FWHM condition, $A(W/2)=1/2$, allows us to determine how $W$ must respond to changes in $J_0$ and $\Delta \equiv b-a$. 

Varying $J_0$ at FWHM we have 
\begin{equation}
\begin{split}
	 \frac{d\log A}{dJ_0} = & \frac{1}{cW/4+J_0} -\frac{1}{b W/2+J_0} \\
      & + \displaystyle\sum_{n=1}^{W/2-1} \left ( \frac{1}{na+J_0} -\frac{1}{nb+J_0}\right ) \\
\end{split}
\end{equation}
where $c/2 < b$  and $a< b$  so  that $d \log A /dJ_0>0$ and $A$ monotonically increases with $J_0$. We conclude that $W$ increases with increasing $J_0$. 

In the same manner,
\begin{equation}
\begin{split}
\frac{d\log A}{d\Delta}  = &  \left( \frac{W}{W \Delta +2aW+4J_0}  - \frac{W}{W \Delta +aW+2J_0}  \right) \\
			  & - \displaystyle\sum_{n=1}^{W/2-1} \frac{n}{n \Delta+n a+J_0} \\
\end{split}
\end{equation}
is negative, so that $A$ monotonically decreases with $\Delta \equiv b-a$. We conclude that $W$ decreases with increasing $b-a$.



\begin{thebibliography}{99}
\bibitem{onestoichiometric} With a pool of $N_g$ monomers and a ratio $t$ of termination rate to monomer addition rate, the length distribution of a 	single ppilus is $p_1(l)=t \Gamma(N_g +1) \Gamma(N_g +t -l)/ [ \Gamma(N_g +t +1) \Gamma(N_g -l +1)]$. For small $t$ this is peaked at  $N_g$.
\end{thebibliography}
\end{document}